\documentclass[useAMS,usenatbib]{mn2e}
\usepackage{times}
\usepackage{graphicx}
\usepackage{color}
\usepackage{ulem}
\usepackage[T1]{fontenc}
\usepackage{aecompl}

\begin{document}

\newcommand{\rma}{R_{\rm MA}}
\newcommand{\stsrma}{(\sigma_{\rm T}/\sigma_{\rm R})_{\rm MA}}
\newcommand{\stsr}{\sigma_{\rm T}/\sigma_{\rm R}}
\newcommand{\rj}{r_{\rm J}}
\newcommand{\rhrj}{r_{\rm h}/r_{\rm J}}
\newcommand{\rtrj}{r_{\rm t}/r_{\rm J}}
\newcommand{\trh}{t_{\rm rh,i}}
\newcommand{\Rh}{R_{\rm h}}
\newcommand{\Nbody}{$N$-body }
\newcommand{\ltorder}{\hbox{ \rlap{\raise 0.425ex\hbox{$<$}}\lower
0.65ex\hbox{$\sim$} }}

\voffset -0.5in

\title[Anisotropy in star clusters]{Velocity anisotropy in tidally limited star clusters}
\author[M. Tiongco et al.]{Maria A. Tiongco$^1$\thanks{E-mail: mtiongco@indiana.edu}, 
Enrico Vesperini$^1$, and Anna Lisa Varri$^2$ \\
$^1$Department of Astronomy, Indiana University, Bloomington, IN 47405, USA \\
$^2$School of Mathematics and Maxwell Institute for Mathematical Sciences, University of Edinburgh, Edinburgh EH9 3JZ, UK}

\maketitle

\begin{abstract}
We explore the long-term evolution of the anisotropy in the velocity space
of star clusters starting with different structural and kinematical properties.  
We show that the evolution of the radial
anisotropy strength and its radial variation within a cluster contain distinct
imprints of the cluster initial structural properties, dynamical
history, and of the external tidal field of its host galaxy.  
Initially isotropic and compact clusters with small initial values of
the ratio of the half-mass to Jacobi radius, $\rhrj$, develop a strong
radial anisotropy during their long-term dynamical evolution.
Many clusters, if formed with small values of $\rhrj$, should now be characterized by a significant  radial
anisotropy increasing with the distance from the cluster centre,
reaching its maximum at a distance between 0.2 $\rj$ and 0.4 $\rj$,
and then becoming more isotropic or mildly
tangentially anisotropic in the outermost regions.  
A similar radial variation of the anisotropy can also result from an
early violent relaxation phase. In both cases, as a cluster continues its evolution  and
loses mass, the anisotropy eventually starts to decrease and the system evolves toward an isotropic
velocity distribution. However, in order to completely erase the strong
anisotropy developed by these compact systems during their evolution,
they must be in the advanced stages of their evolution and lose
 a large fraction of their initial mass. 
Clusters that are initially isotropic and characterized by larger
initial values of $\rhrj$, on the other hand, never develop a significant radial anisotropy. 

\end{abstract}

\section{Introduction}
A complete characterization of globular cluster dynamical properties
requires information on both the cluster structure and
kinematics. Observational studies  have extensively explored the
internal spatial structure of Galactic clusters \citep[see e.g.][]{djorgovski1994, mclaughlin2005, miocchi2013} and
significant efforts have been invested in  theoretical
investigations aimed at understanding the cluster structural evolution
and at providing numerical and analytical models for the interpretation
of observational data (see e.g. \citealt{heggie2003} and references therein). 

The observational study of the internal
kinematical properties of globular clusters, on the other hand, is
much more challenging. After a few early efforts (see e.g. \citealt{gunn1979,lupton1987,pryor1993, gebhardt1995, vanleeuwen2000}; see also \citealt{meylan1997} and references therein),
only recently observational investigations have started to provide
more extensive information on globular cluster kinematics thanks to, for example, ESO/VLT
radial velocity \citep[see e.g.][]{bellazzini2012, lanzoni2013, lardo2015} and HST proper motion measurements \citep{bellini2014, watkins2015}. Forthcoming measurements from the Gaia astrometric mission
will soon provide a wealth of new data and further enrich the
observational study of cluster kinematics \citep[see e.g.][]{pancino2013, sollima2015}. 

A more detailed knowledge of  clusters' kinematics allows to address
a number of fundamental questions concerning their
dynamics and stellar content  such as, for example, the strength of internal
rotation, the possible link between internal rotation and cluster morphology \citep[see e.g.][]{bianchini2013, fabricius2014,kacharov2014}, the presence of intermediate-mass black holes \citep[see e.g.][]{lutzgendorf2011, anderson2010, lanzoni2013}, and the possible differences in the dynamical history of the
multiple stellar populations observed in many clusters \citep[][]{richer2013, bellini2015}. 

In this paper we present the results of a survey of \Nbody
simulations following the long-term evolution of systems with a broad
range of different initial structural and kinematical properties and
including the effects of an external tidal field. Our study is
focused on the evolution of the  anisotropy in the velocity
space and aimed at exploring the connection between the
evolution of the velocity anisotropy and the cluster initial
conditions  and dynamical history. We will show that different
initial structural properties, different evolutionary phases, 
and the external tidal field leave distinct imprints
on the strength of the velocity distribution anisotropy and its radial
variation within a cluster. The results obtained further emphasize the wealth of crucial dynamical information contained in the kinematical properties and the importance of pursuing observational studies of cluster internal kinematics.
The outline of the paper is the following: in Section 2 we describe
method and initial conditions, in Section 3  we present our results,
and in Section 4 we summarize our conclusions.

\section{Method and Initial Conditions}
\label{sec:method}

\begin{table}
\caption{Simulation Information.}
\label{tab:details}
\begin{tabular}{@{}llllllll}
\hline
Model ID & 
$(\rhrj)_{\rm i}$ & 
$(\rtrj)_{\rm i}$\\
\hline
KF1 & 0.116 & 1.0\\
KF075 & 0.087 & 0.75\\
KF05 & 0.058 & 0.5\\
KF025 & 0.029 & 0.25\\
KF0125 & 0.012 & 0.125\\
\\
VBrotF1 & 0.232 & 1.0\\
VBrotF09 & 0.209 & 0.9\\
VBrotF075 & 0.174 & 0.75\\
VBrotF05 & 0.116 & 0.5\\
VBrotF04 & 0.093 & 0.4\\
VBrotF025 & 0.058 & 0.25\\
VBrotF0125 & 0.029 & 0.125\\
VBrotF005 & 0.012 & 0.05\\
\\
vrQ01F05 & 0.095 & -\\
vrQ01F02 & 0.036 & -\\
vrQ001F05 & 0.036 & -\\
\hline
\end{tabular}
\end{table}

The $N$-body simulations presented in this paper were carried out using the {\sevensize NBODY6} code \citep{aarseth2003} accelerated by a GPU \citep{nitadori2012}, and run on the {\sevensize BIG RED II} cluster at Indiana University.

Clusters are assumed to be on circular orbits in the host galaxy tidal
field modeled as a point-mass, and the equations of motion are solved in a frame of reference co-rotating with the cluster around the host galaxy centre (see e.g. \citealt{heggie2003}).  For all the simulations we consider
systems with $N=16,384$ equal-mass particles; particles moving beyond
two times the Jacobi radius, $\rj$, are removed from the simulation.  All the simulations are run until half of the initial cluster
mass is lost; we have run 4 different random realizations of the initial conditions of each of the models to provide an indication of the extent of stochastic variations in the results.

For the initial conditions we consider three sets of models
characterized by different structural and kinematical properties.
The first set of simulations follow the evolution of isotropic King
models \citeyearpar{king1966} with 
dimensionless central potential $W_0=7$ and different values of the
filling factor, defined as the ratio of the half-mass radius to the
Jacobi radius $\rhrj$ or equivalently, the ratio of the model truncation radius to the Jacobi radius $\rtrj$ (see Table \ref{tab:details}).

For the second set of models we have considered a series of rotating initial conditions, sampled from the equilibria introduced by \citet{varri2012}. This family of distribution function-based models was specifically designed to describe quasi-relaxed stellar systems with finite total angular momentum, and the resulting configurations are characterized by differential rotation, approximately rigid in the centre and vanishing in the outer parts, and by pressure anisotropy. Specifically, as a result of the chosen truncation prescription of the distribution function in phase space and the requirement of self-consistency of the associated density-potential pair, the velocity dispersion tensor of the models is characterized by isotropy in the central region, weak radial anisotropy in the intermediate regions, and tangential anisotropy in the outer parts. For a complete description we refer the reader to \citet{varri2012}, here we just recall that the models are defined by four dimensionless parameters, with the concentration parameter $W_0$ (defined as the depth of the dimensionless potential well at the centre of the system), and a parameter $\hat{\omega}$ measuring the strength of the rotation, playing the leading roles. The additional parameters $\bar{b}$ and $c$ are determined by the analytical expression of the distribution function, and they  determine the shape of the rotation radial profile.

The rotating models selected for this investigation are characterized by the following values of the dimensionless parameters: $W_0 = 6, \hat{\omega} = 0.25, \bar{b} = 0.05,$ and $c =1$. Such a choice is inspired by results of a first comparison between the \citet{varri2012} models and selected Galactic globular clusters (the properties of the best-fit models of 47 Tuc, $\omega$ Cen, and M15, are available in Tables 3, 4 of \citealt{bianchini2013}). The coupling between the angular momentum vectors associated with the internal rotation and the orbital motion is a non-trivial aspect of the definition of our initial conditions. For simplicity, in our current set-up the rotation axis is set to be perpendicular to the orbital plane and  the cluster internal rotation is prograde (with respect to the orbital motion). As in the previous set of simulations, also in this case we have explored several different values of the initial filling factor (see Table \ref{tab:details}). The dynamical interplay between internal rotation and external tidal field is a very rich field of investigation, which, so far, has been only partially explored \citep[see e.g.][]{ernst2007,hong2013}. Indeed, this set of simulations is only a part of an extended survey aimed at studying the role of rotation in the long-term dynamical evolution of tidally limited rotating star clusters (Tiongco et al., in prep.).

The third set of simulations follow the evolution of models initially
undergoing a phase of violent relaxation in the host galaxy external tidal field with different initial values
of the virial ratio, $Q = T/|V|$ (where $T$ and $V$ are the system
kinetic energy and potential energy respectively), and different
ratios of the initial cluster limiting radius to the Jacobi radius
($r_{\rm L}/\rj$).  We use the following parameters for the initial conditions of the violent relaxation phase: (ID, $Q$, $r_{\rm L}/\rj$): (vrQ01F05, 0.1, 0.5),  (vrQ01F02, 0.1, 0.2), and (vrQ001F05, 0.01, 0.5).  The systems are initially spherical and
homogeneous. During the violent relaxation phase, the models develop
an isotropic core, radial anisotropy in the intermediate regions, and
either isotropic or tangentially anisotropic outer regions.  The
kinematical properties emerging at the end of the  violent
relaxation are qualitatively similar to those discussed in
\cite{vesperini2014}; in this paper we focus on the
\textit{long-term} evolution of these systems.
For these simulations the properties reported in Table
\ref{tab:details} are those of the equilibrium systems at the end of
the violent relaxation phase (averages of 4 realizations). We refer to the equilibrium properties of these systems as our initial conditions for the long-term evolution. These simulations were started with a slightly larger number of particles in order to take into account of the early loss of stars during the violent relaxation phase. At the end of the violent relaxation phase the vrQ01F05, vrQ01F02, and the vrQ001F05 systems had, respectively, 14,860, 16,297, and 17,091 particles (averages of 4 realizations).

To measure the anisotropy we use the ratio of the projected
tangential to the radial velocity dispersion, $\stsr$, measured on the cluster orbital plane. For the calculation of the velocity dispersions and their radial variation within the cluster, cylindrical shells along a line of sight perpendicular to the orbital plane with heights encompassing the entire cluster and at different projected radii, $R$, have been used.  We have combined 5 snapshots around the desired times for each of the 4 realizations, then taking the profiles from each realization at the same times, calculated the median $\stsr$ at each $R$ to get the median profile.  We similarly find the upper and lower bounding profiles to represent the radial variation of the range of $\stsr$.  In the construction of the profiles all particles enclosed within the Jacobi radius have been taken into account.

\section{Results}

\subsection{Evolution of the velocity anisotropy}
\label{sec:sec31}

\begin{figure}
\centering
\includegraphics[width=3.3in]{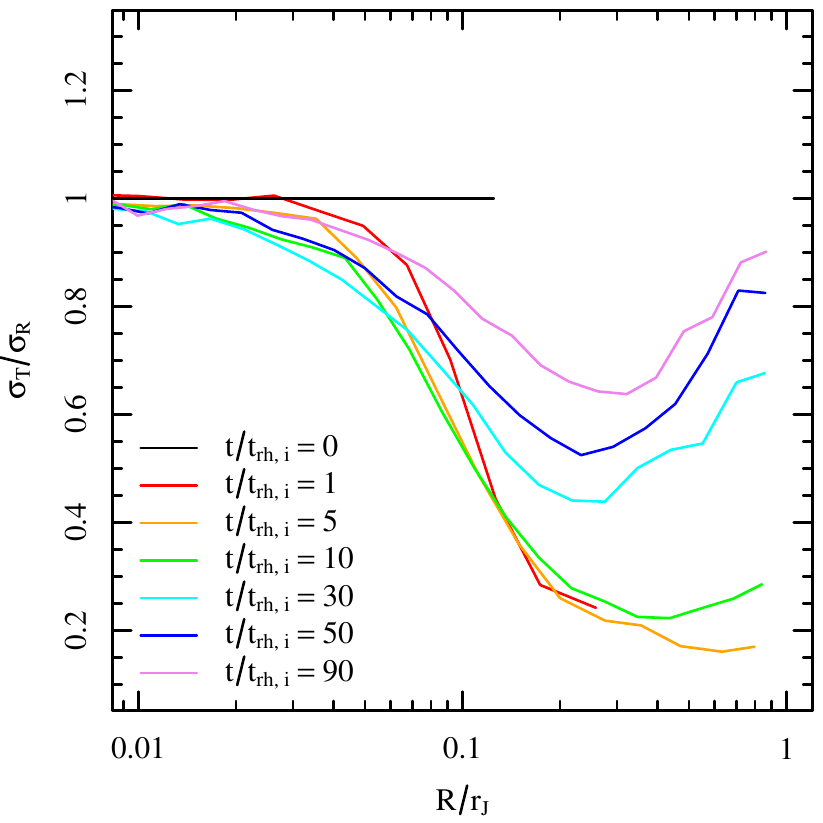}
\caption{Time evolution of the radial profile of $\stsr$ for the model KF0125.  The radius is normalized to the Jacobi radius, $\rj$. Each line is the median of the profiles obtained from the 4 realizations; see Section \ref{sec:method} for details. The theoretical isotropic profile of the King model at $t=0$ is shown.}
\label{fig:evol_king}
\end{figure}

\begin{figure}
\centering
\includegraphics[width=3.3in]{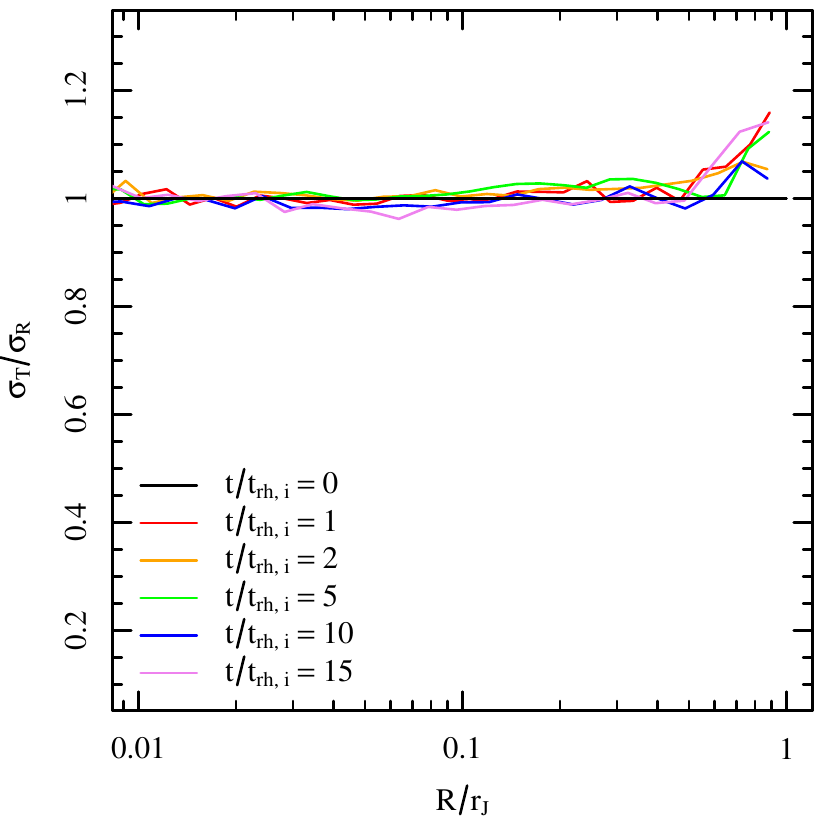}
\caption{Same as Figure \ref{fig:evol_king} for the model KF1.}
\label{fig:evol_kingF1}
\end{figure}

\begin{figure}
\centering
\includegraphics[width=3.3in]{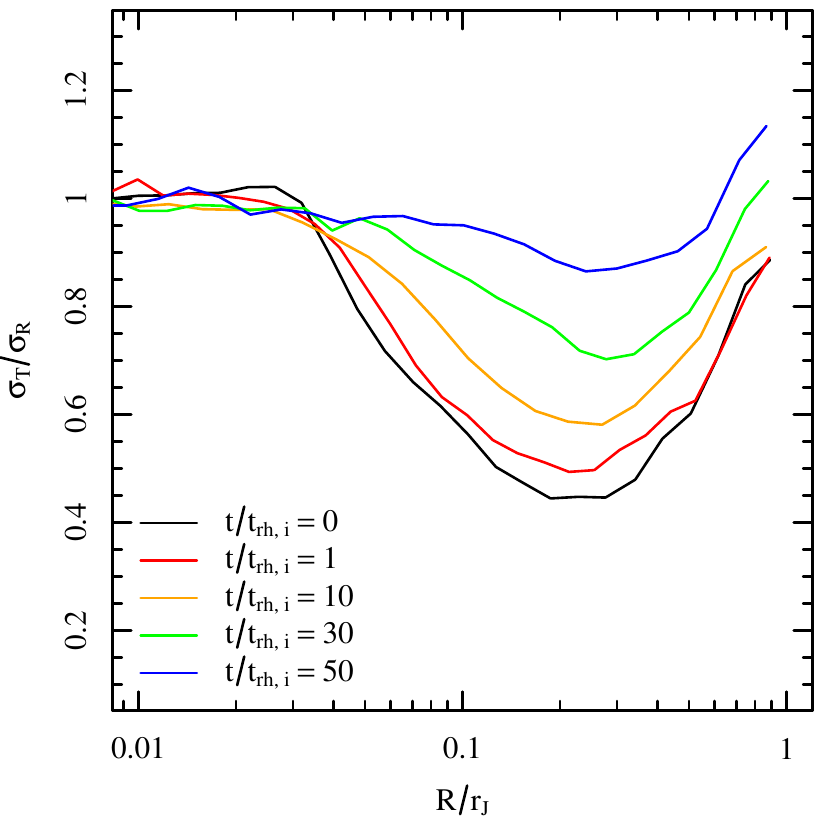}
\caption{Same as Figure \ref{fig:evol_king} for the model vrQ01F02.  }
\label{fig:evol_vr}
\end{figure}

\begin{figure}
\centering
\includegraphics[width=3.3in]{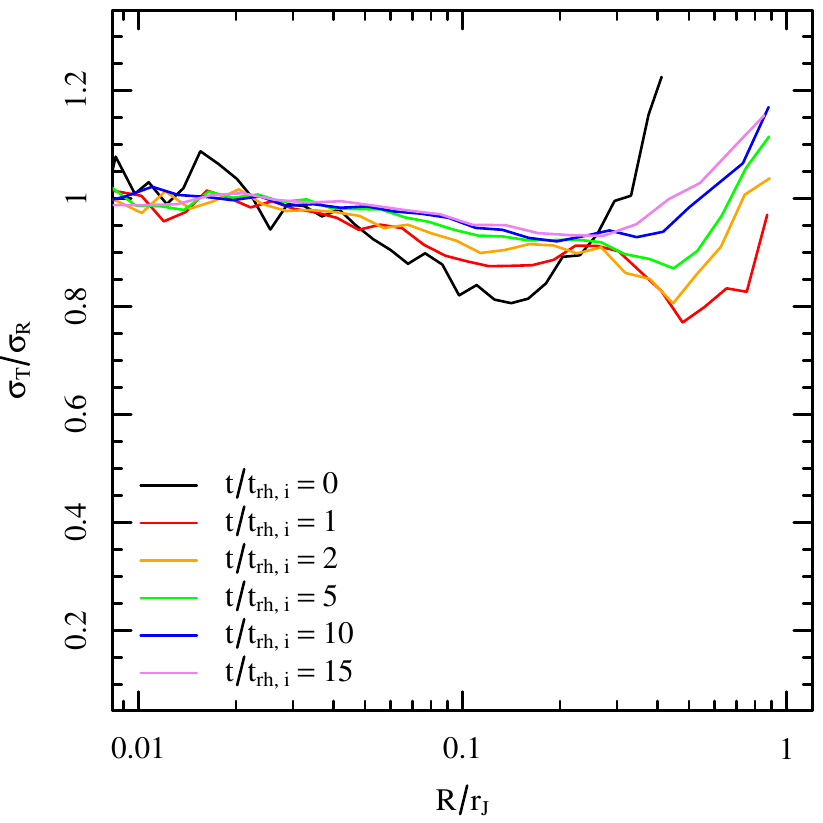}
\caption{Same as  Figure \ref{fig:evol_king} for model VBrotF05.}
\label{fig:evol_rot}
\end{figure}

As shown in a number of studies \citep[see e.g.][]{giersz1994,giersz1994b,takahashi1995,takahashi1996,takahashi1997,spurzem1996}, during their
long-term evolution initially isotropic and isolated star clusters
develop a strong outer radial anisotropy. This feature may be easily 
interpreted as the orbital distribution in the halo being dominated by radial orbits, as a result of two-body relaxation in the cluster central regions.
For tidally limited systems, on the other hand, the cluster structural evolution
and expansion is limited by the tidal boundary, the growth of the
radial anisotropy is suppressed by the preferential loss of stars on
radial orbits resulting in an isotropic (or tangentially anisotropic) velocity
distribution  in the cluster outermost regions
\citep[see e.g.][]{takahashietal1997,takahashi2000,baumgardt2003,hurley2012}.  As a
cluster loses mass, its Jacobi radius will move toward the inner
regions where the
velocity dispersions are more 
isotropic, erasing the anisotropy that might have developed in the
cluster intermediate regions
\citep[see][]{giersz1997}.  

We set the ground by examining in detail the time evolution of the radial anisotropy profile of two representative cases of the first set of simulations, i.e., the models starting from isotropic, non-rotating, King (1966) initial conditions. First, we consider the most underfilling case, with $(\rtrj)_{\rm i}=0.125$ (KF0125). As illustrated in Figure 1, the cluster is initially isotropic and underfilling, and, as it evolves and expands, it develops a strong radial anisotropy in the outer regions.  As the system continues its evolution and starts to lose mass, a
minimum in the anisotropy profile forms (corresponding to a maximum in the radial anisotropy) while the outermost regions become increasingly
less radially anisotropic. This behavior is indeed consistent with the results of previous investigations based on \Nbody, Fokker-Planck, and gaseous methods discussed at the beginning of this section.  A similar radial variation in the velocity anisotropy is found in models based on the Michie-King distribution function (\citealt{michie1963}; see also \citealt{gieles2015} for a recent study presenting a new family of anisotropic models).  After this phase of the evolution, the overall strength of the radial anisotropy and the depth of the minimum will decrease.  We will discuss in detail the evolution of the maximum radial anisotropy, its magnitude and location relative to the Jacobi radius over time as well as the dependence on the cluster initial conditions in Section \ref{minimum}.

We then continue the analysis of the first set of simulations by
considering the tidally filling case, i.e. such that ($\rtrj)_{\rm i}=1$ (KF1).
As depicted in Figure \ref{fig:evol_kingF1}, such a tidally filling
system never develops a significant degree of radial anisotropy,
especially the outermost parts, which are instead characterized by (mild)
tangential anisotropy. Such a behavior is consistent with the results
of previous investigations by \citet{baumgardt2003}. We emphasize
that the development of such tangentiality is quite rapid 
(i.e., $t \sim$ 1 $\trh$), 
and it has been interpreted as a result of a preferential 
loss of stars on radial orbits in the outer parts of the systems,
although the details of the processes underpinning such a behaviour
are still only partly understood \citep[see e.g.][]{keenan1975,oh1992}.  Further work to clarify the role of the loss of stars and of the external tidal field in the development of tangentiality is necessary.
                               
A comparison between these two extreme cases clearly illustrates the
importance of the initial filling factor of a system, and envisages
the existence of a fundamental connection between a cluster
kinematics and its initial properties and dynamical history.  We will
further discuss the implications of these differing kinematical
signatures for clusters with different initial filling factors in
Section \ref{minimum}.   

It is now particularly instructive to move on to the third set of
simulations, exploring the evolution of the products of violent
relaxation in an external tidal field. Here our motivation was
specifically to evaluate the long-term effects of the tidal
perturbation, which was already responsible for having shaped the
kinematic and structural properties on the systems in the early phase
of their dynamical evolution (for details, see \citealt{vesperini2014}). As a representative case, in Figure \ref{fig:evol_vr} we examine model
vrQ01F02, which is characterized by $Q=0.1$ and $r_{\rm L}/\rj = 0.2$,
corresponding to intermediate values within the ranges explored in the
third set of simulations.  

Indeed, systems emerging from the violent relaxation phase start their
long-term evolution with an anisotropy profile qualitatively similar
to that produced on a much longer timescale in the underfilling
KF0125 model (see Figure \ref{fig:evol_vr}, black line corresponding to $t/\trh = 0$), and
the subsequent evolution is remarkably self-similar, driving the
system toward a state characterized by a more isotropic orbital
distribution (see Figure \ref{fig:evol_vr}, blue line corresponding to
$t/\trh=50$). Such an evolution is reflected in the decrease of the
magnitude of the maximum of the radially-biased portion of the
profile. The extent of the maximum radial anisotropy depends on the
initial filling factor and virial ratio with colder and/or more
underfilling initial conditions producing a stronger radial anisotropy
\citep[see also][]{vesperini2014}. 

Finally, we now consider the systems initially characterized by internal rotation, investigated in the second set of simulations. For our detailed analysis, illustrated in Figure \ref{fig:evol_rot}, we focused on model VBrotF05 which, again, corresponds to an intermediate filling factor, among the values considered in this section of our investigation.

The initial conditions (see Figure \ref{fig:evol_rot}, black line corresponding to
$t/\trh = 0$) have an isotropic core, a radially anisotropic intermediate
region, and a tangentially isotropic outer region. We emphasize that, in this case, the initial radial anisotropy and the location of its
maximum are {\it intrinsic} properties of the model adopted and not the result of the cluster early or late
evolution. The location of the maximum anisotropy in the initial model is near the half-mass radius, and its location relative to the Jacobi radius will depend on the initial filling factor.
  We also point out that the strength of the intrinsic projected anisotropy of the rotating model depends on the angle between the line of sight and the rotation axis (it is larger when the line of sight is parallel to the rotation axis and decreases as the line of sight moves to a direction perpendicular the rotation axis).
  The shape of the anisotropy profile during the
first 1-2 relaxation times will also depend on the initial filling factor.  
During the evolution
of underfilling VBrot systems, the intrinsic anisotropy is erased
while the system is expanding and its outer anisotropy  evolution
resembles that shown in Figure \ref{fig:evol_king} for initially underfillling King models (see Figure \ref{fig:evol_rot}, lines corresponding to $t/\trh=10, 15$).  
Underfilling models may be characterized by two minima in the profile at early
times, the inner one being intrinsic to the model, and the outer one
resulting from the cluster evolution (see the red line corresponding to $t/\trh=1$ in Figure  \ref{fig:evol_rot}).  
For the more filling VBrot systems, the model's intrinsic radial
anisotropy is initially 
located near the range of radial distances where the anisotropy
produced by evolution develops. Similarly to what happens for the more
filling King models, these models develop only a mild tangential
anisotropy in the outer regions during their evolution. 

\subsection{Properties of the maximum radial velocity anisotropy}
\label{minimum}

\begin{figure}
\centering
\includegraphics[width=3.3in,height=5in]{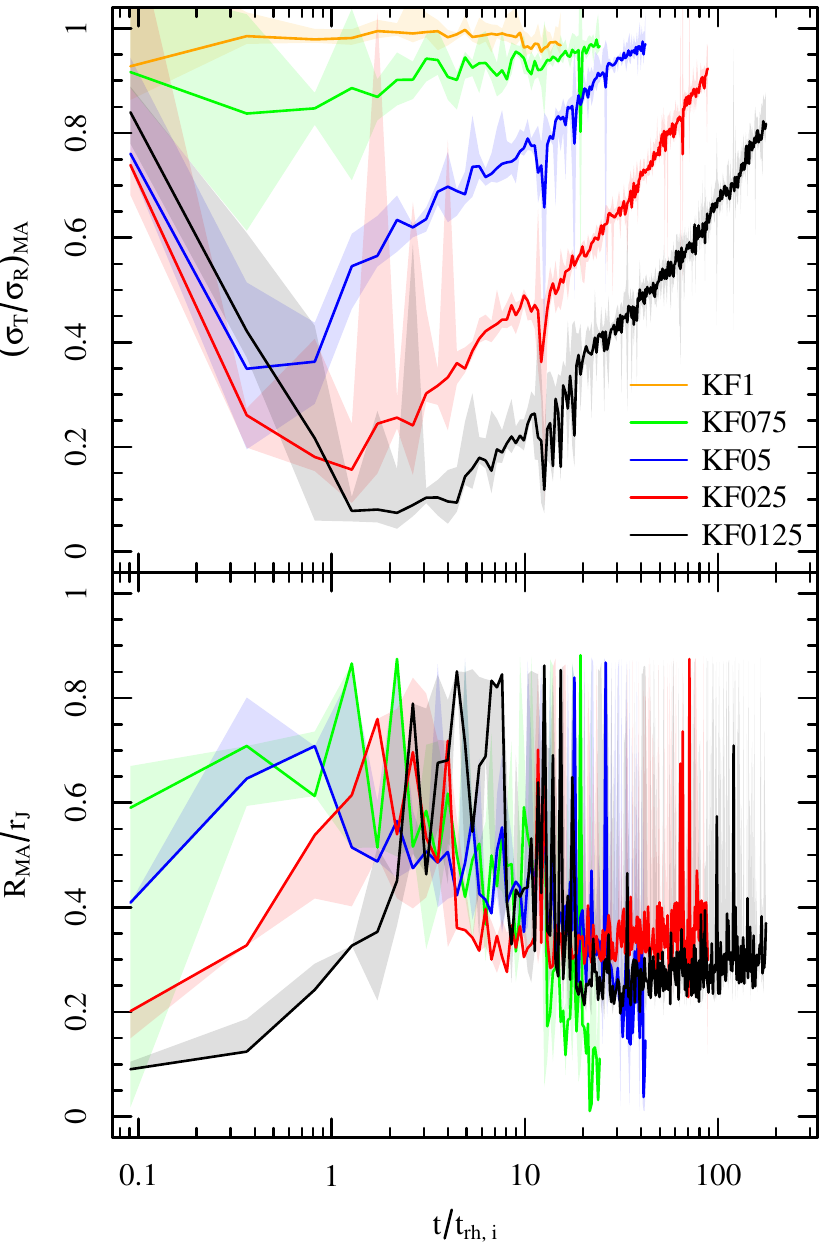}
\caption{Top Panel: Time evolution of the value of the maximum radial anisotropy for the isotropic King models.  The solid line is, for each time, the maximum anisotropy of the median profile (see Section \ref{sec:method} for explanation and Figures \ref{fig:evol_king}-\ref{fig:evol_rot} for examples), while the upper and lower bounds represent the range of the anisotropy of the 4 realizations at that location. Bottom panel: Evolution of the location of the maximum radial anisotropy relative to the Jacobi radius. The solid line, upper, and lower bounds represent the location of the maximum anisotropy for the median, upper, and lower bounding profiles at each time for the 4 simulations.  The line for KF1 is omitted from the bottom panel because this model never develops a significant radial anisotropy.}
\label{fig:ma_king}
\end{figure}

\begin{figure}
\centering
\includegraphics[width=3.3in,height=5in]{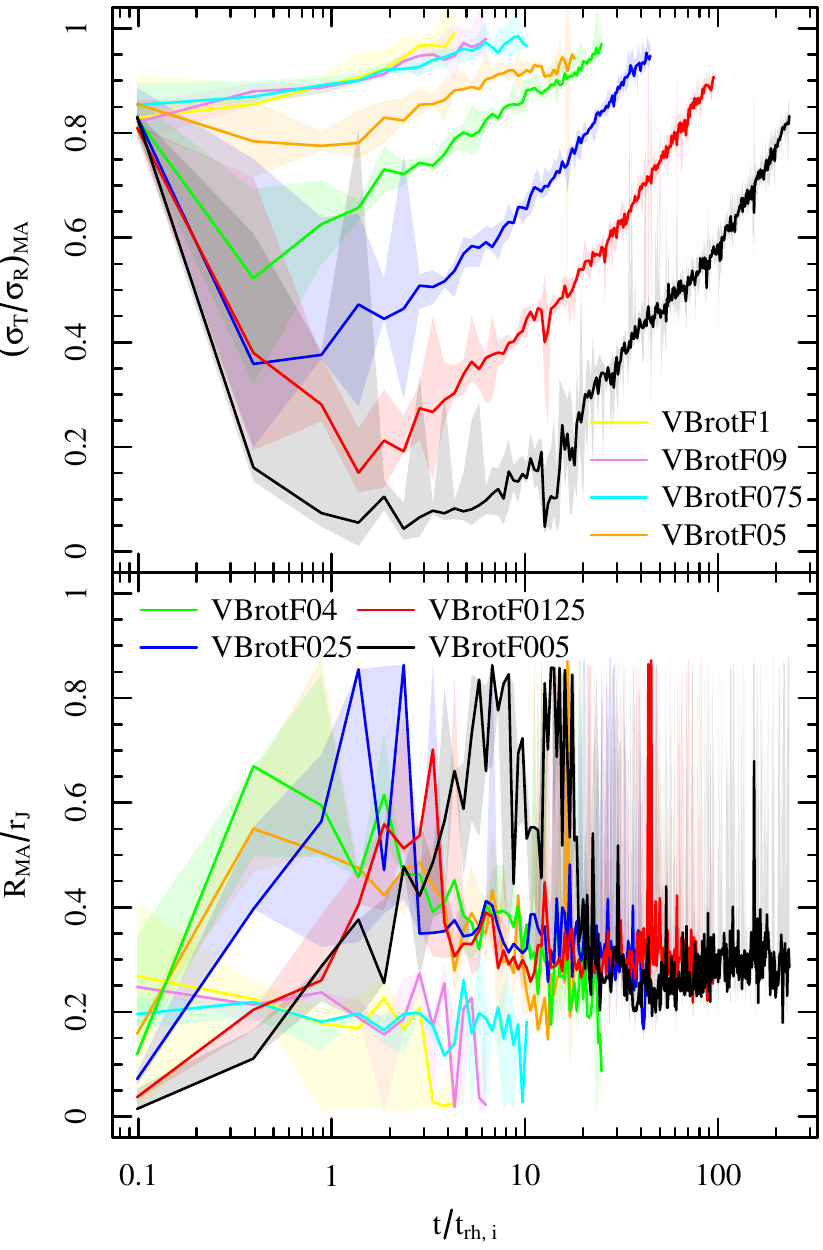}
\caption{Same as Figure \ref{fig:ma_king} for the VBrot models.}
\label{fig:ma_rot}
\end{figure}

\begin{figure}
\centering
\includegraphics[width=3.3in,height=5in]{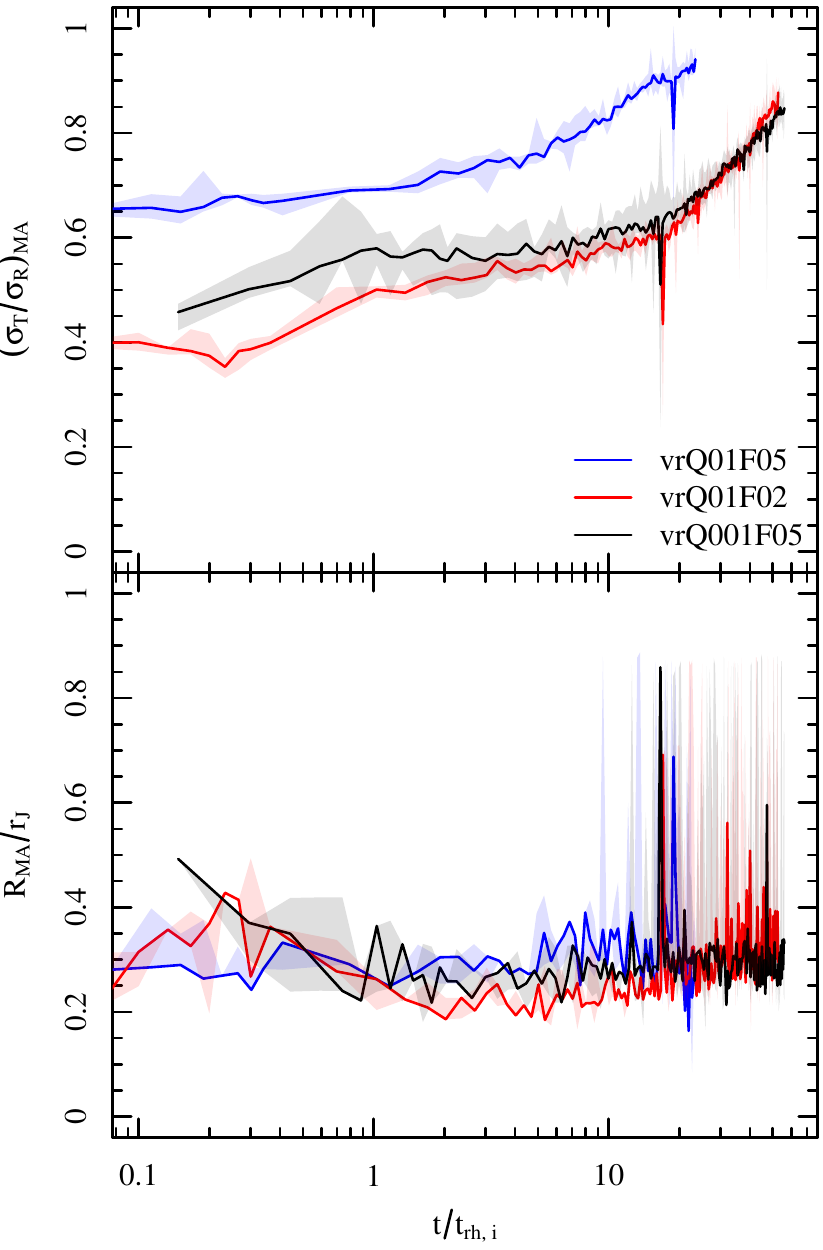}
\caption{Same as Figure \ref{fig:ma_king} for the vr models.}
\label{fig:ma_vr}
\end{figure}

\begin{figure}
\centering
\includegraphics[width=3.3in]{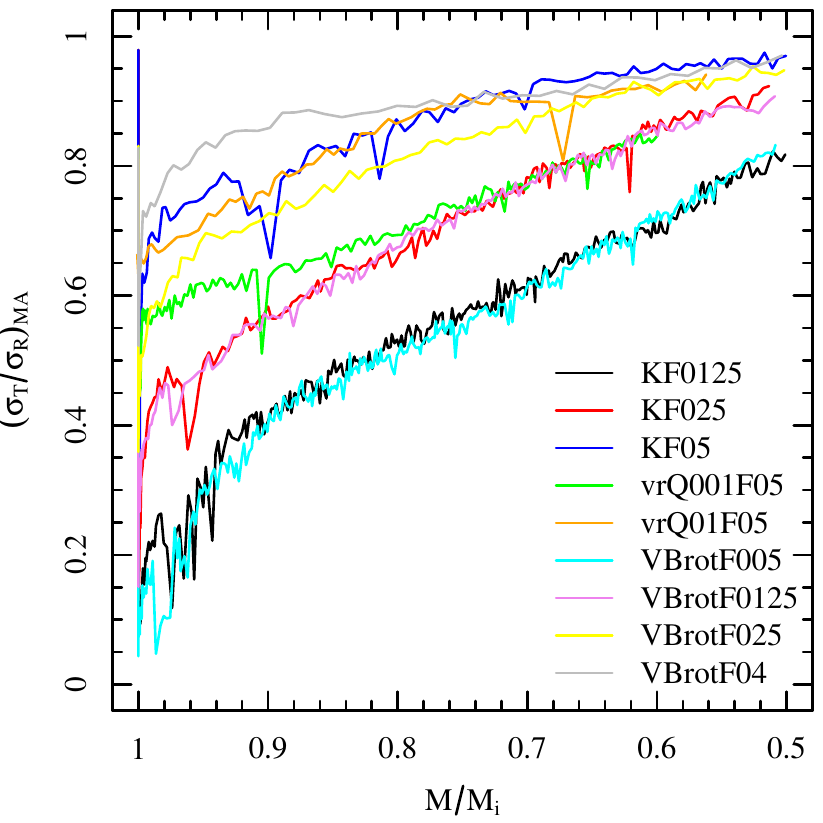}
\caption{Evolution of the value of the maximum radial anisotropy as function of fraction of initial mass remaining for several models.}
\label{fig:ma_mass}
\end{figure}

\begin{figure}
\centering
\includegraphics[width=3.3in]{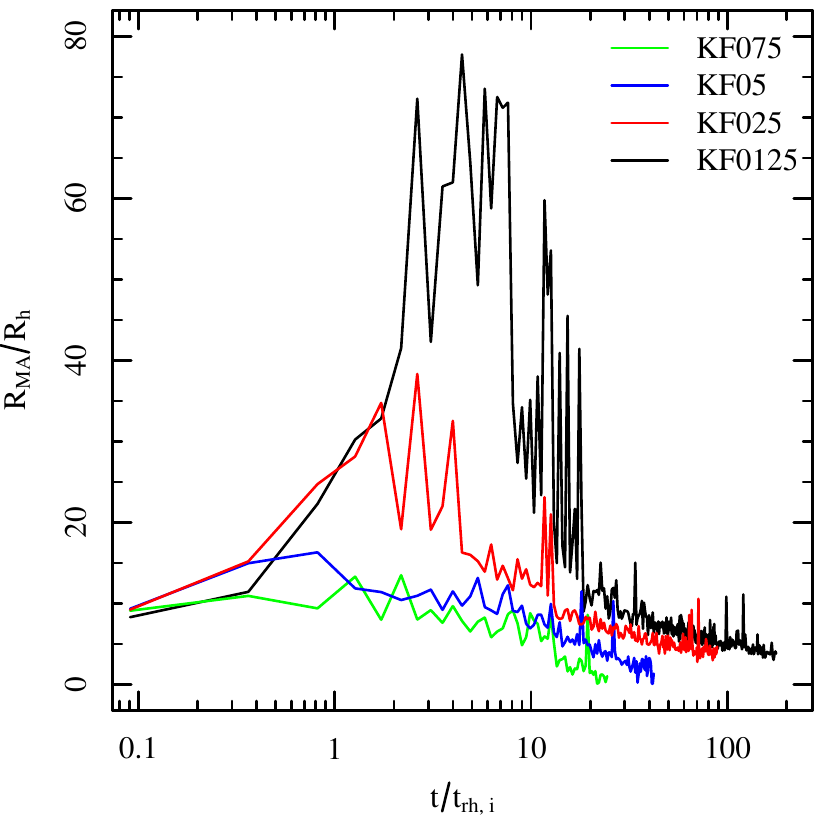}
\caption{Evolution of $\rma$ in terms of the projected half-mass radius, $\Rh$ for the King models.}
\label{fig:ma_Rh}
\end{figure}

In the previous section we have illustrated how differences in a cluster initial conditions and evolutionary phases can affect the evolution of a cluster kinematics and leave different fingerprints in the anisotropy radial profile. In this section, we further explore the connection between initial conditions, dynamical evolution and radial anisotropy by focusing our attention on the strength of the maximum radial anisotropy, $\stsrma$, and its location, $\rma$.
We show in Figures \ref{fig:ma_king}-\ref{fig:ma_vr} the time evolution 
$\stsrma$ and $\rma$, relative to the Jacobi radius
for each set of models.  More underfilling initial conditions develop stronger radial anisotropy while, for all systems, the location of 
the maximum radial anisotropy is between 0.2 and 0.4 $r_J$ for most of the cluster's lifetime.   

As for the time evolution of $\rma$ we point out that, as was shown in Figure \ref{fig:evol_king}, the rising portion of the anisotropy profile appears only after the system expanded to the outermost regions approaching the Jacobi radius.
The development of this feature in the anisotropy radial profile
corresponds to the time when $\rma/\rj$ stops increasing and $\stsrma$ stops decreasing
(see Figure \ref{fig:ma_king}). Figure \ref{fig:ma_king} clearly shows
the connection between a cluster initial filling factor and the
strength  of the radial anisotropy developed during its dynamical
evolution: initially compact clusters with smaller initial filling
factors develop stronger radial anisotropy. Only during the late
stages of their evolution when they have lost a significant fraction
of their mass, underfilling systems evolve toward a more isotropic
velocity distribution (we emphasize that we refer here to mass loss
occurring during the cluster long-term evolution and due to two-body
relaxation not to early mass loss due to stellar evolution). This point is further illustrated in Figure \ref{fig:ma_mass}
which shows the evolution of the maximum radial anisotropy as a function of mass lost for several underfilling models. 
Underfilling systems develop a strong radial anisotropy which is gradually erased as
the cluster evolves and loses mass. Our simulations show that clusters
which are initially significantly underfilling  should now be
characterized in their intermediate regions by a strong radial anisotropy unless they are in the very
advanced stages of their evolution and have lost a significantly
fraction of their initial mass. Many clusters should not have suffered such a strong relaxation-driven mass loss and information on their initial structural properties should therefore be imprinted in the strength of the radial anisotropy in their outer regions.

Figure \ref{fig:ma_rot} shows the time evolution of $\rma$ and
$\stsrma$ for the VBrot models. As discussed in Section \ref{sec:method}
these models are characterized by an initial intrinsic maximum in the
radial anisotropy. Therefore, the initial 
radial location of the maximum anisotropy relative to the Jacobi radius depends on the initial
filling factor and, in general, does not fall in the range of values
($\rma/\rj \sim 0.2-0.4$) typical for the maximum anisotropy
produced by early or long-term dynamical evolution. Only for the
initially more filling models the intrinsic maximum radial anisotropy
is at 0.2-0.3 $r_J$. For underfilling models, the maximum radial anisotropy rapidly (within 1 $\trh$) transitions from the intrinsic maximum to the one developed in the outermost regions of the cluster as it expands (see for example the radial profile of $\stsr$ at $t/\trh=1$ in Figure \ref{fig:evol_rot}).  For the most filling models
(VBrotF075, VBrotF09, and VBrotF1) dynamical evolution does not
produce a significant radial anistropy; for these models, Figure \ref{fig:ma_rot} shows the gradual erasing of the intrinsic anistropy. 

The vr models also begin their long-term evolution with a velocity
distribution characterized by a region of significant radial anisotropy produced early during the cluster violent relaxation phase  and  the
radial position of the maximum radial anisotropy falls between 0.2-0.4 $\rj$. The evolution of $\stsrma$ and $\rma$ for these models is similar to that found for the King models after the development of a minimum in the $\stsr$ radial profile.  Figure \ref{fig:ma_vr} shows that for the vr models the maximum radial anisotropy will decrease over time while $\rma$ will generally remain between 0.2 and 0.4 $\rj$.

\subsection{Evolution of the anisotropy at $\Rh$}

\begin{figure}
\centering
\includegraphics[width=3.3in]{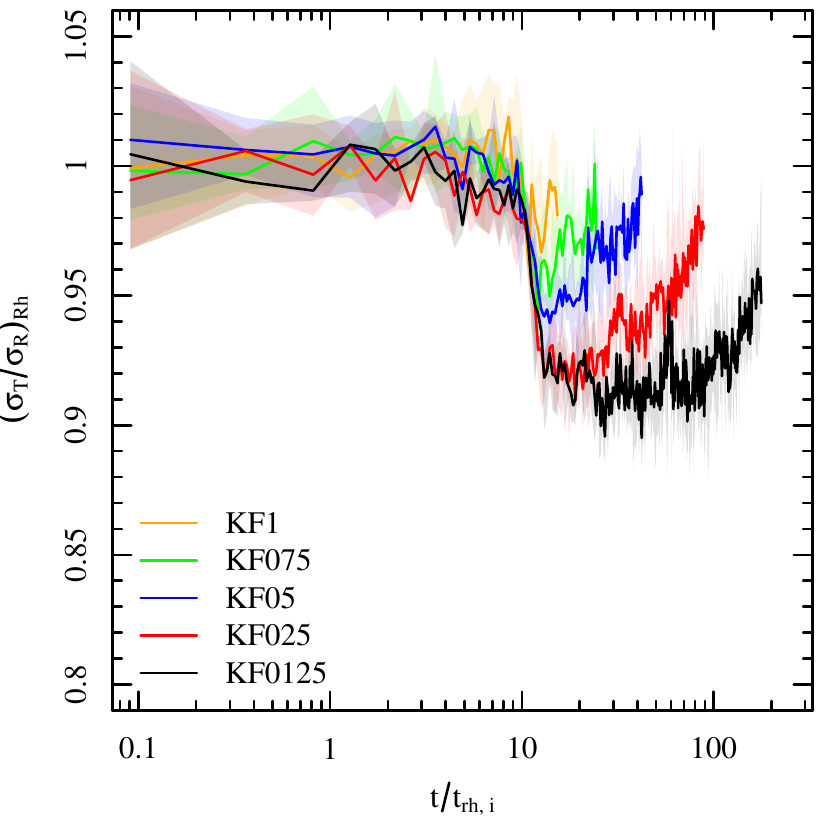}
\caption{Evolution of the anisotropy at $\Rh$ for the King models.}
\label{fig:Rh_king}
\end{figure}

\begin{figure}
\centering
\includegraphics[width=3.3in]{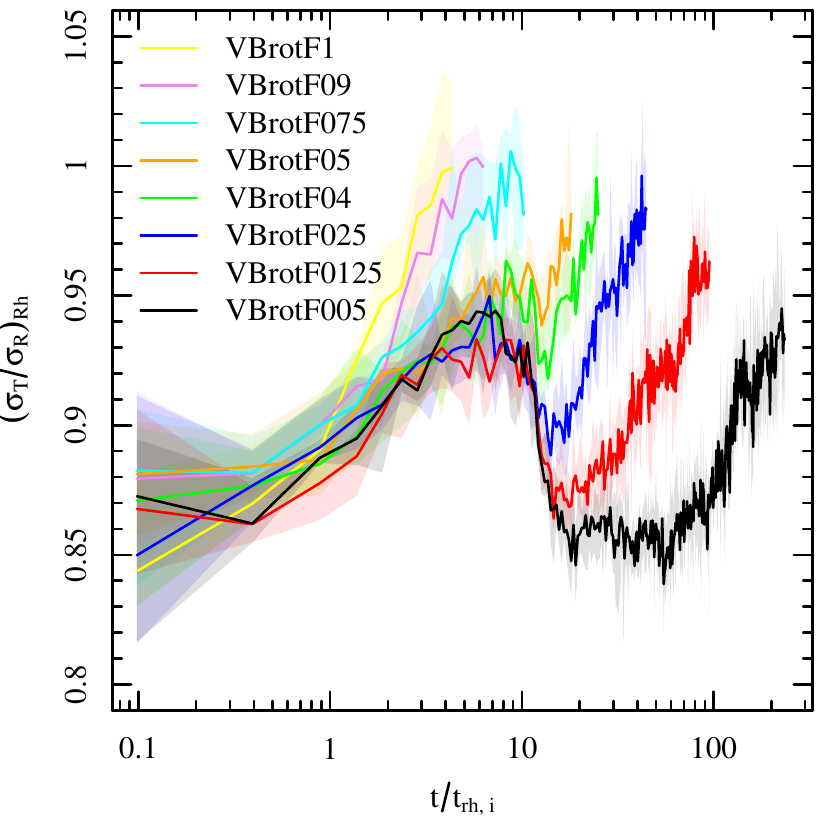}
\caption{Evolution of the anisotropy at $\Rh$ for the VBrot models.}
\label{fig:Rh_rot}
\end{figure}

\begin{figure}
\centering
\includegraphics[width=3.3in]{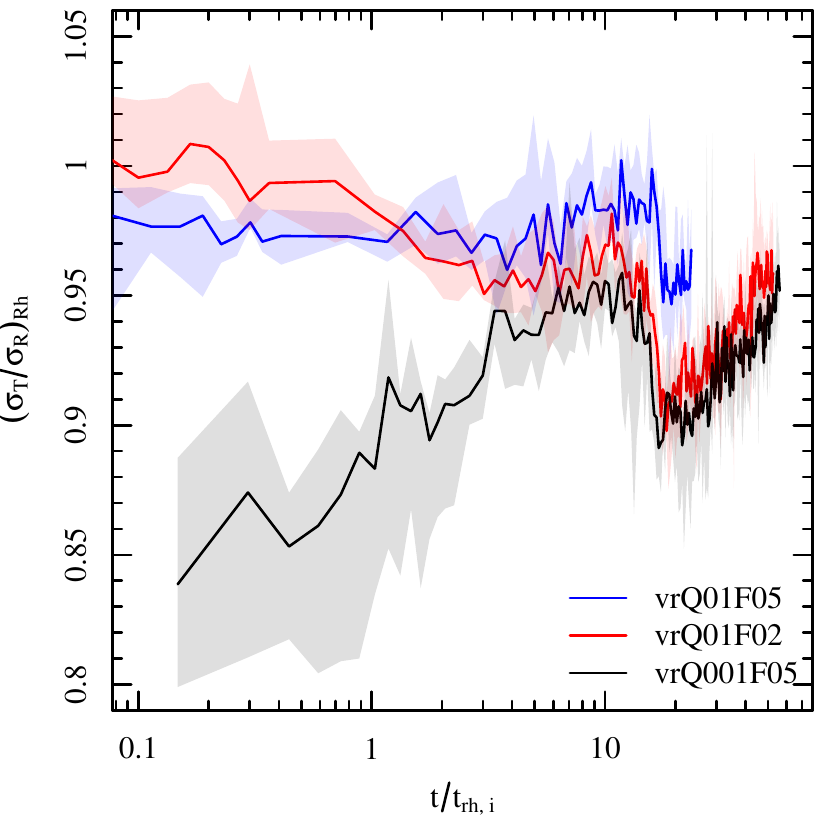}
\caption{Evolution of the anisotropy at $\Rh$ for the vr models.}
\label{fig:Rh_vr}
\end{figure}

The results presented in the previous section show that for the models
developing a significant radial anisotropy the location of the maximum
radial anisotropy falls in the range 0.2-0.4 $\rj$ for most of a
cluster's lifetime; as shown in Figure \ref{fig:ma_Rh}, this range can
correspond to a distance of many half-mass radii from the cluster
centre.  On the observational side, anisotropy measurements in
globular clusters do not cover, in most cases,  distances larger than
about 1-2 half-mass radii, and the deviations from isotropy found are
modest (see e.g. the results of the Hubble Space Telescope Proper
Motion (HSTPROMO) catalogs of Galactic Globular Clusters:
\citealt{bellini2014} and \citealt{watkins2015}, where the values of
$\stsr$ at the half-mass radius of radially anisotropic clusters are
equal to about 0.9-0.95 for the most radially anisotropic systems).

Motivated by these observational constraints, we have also evaluated
the time evolution of the anisotropy measured at $\Rh$ for all models
of our survey. As illustrated in Figures
\ref{fig:Rh_king}-\ref{fig:Rh_vr}, models 
that are initially radially anisotropic (i.e., VBrot and vr models)
become more isotropic over time, although there is a short time
interval characterized by a rapid increase in radial anisotropy at the
time of core collapse. After core collapse, all the systems investigated evolve again towards isotropy. For the VBrot models the anisotropy in the initial phase of evolution ($t \ltorder$ 2 $\trh$) is that associated to the intrinsic anisotropy of the rotating models (as already pointed out in Section \ref{sec:sec31}, the strength of the intrinsic projected anisotropy for the VBrot models depends on the angle between the line of sight and the rotation axis. Figure \ref{fig:Rh_rot} shows the values measured along a line of sight parallel to the rotation axis; for a  45 (90) degree angle, for $t \ltorder$ 2 $\trh$, $\stsr$ at $\Rh$ is equal to about 0.95 (1)).  

\begin{figure*}
\begin{minipage}{6.6in}
\centering
\includegraphics[width=6.6in]{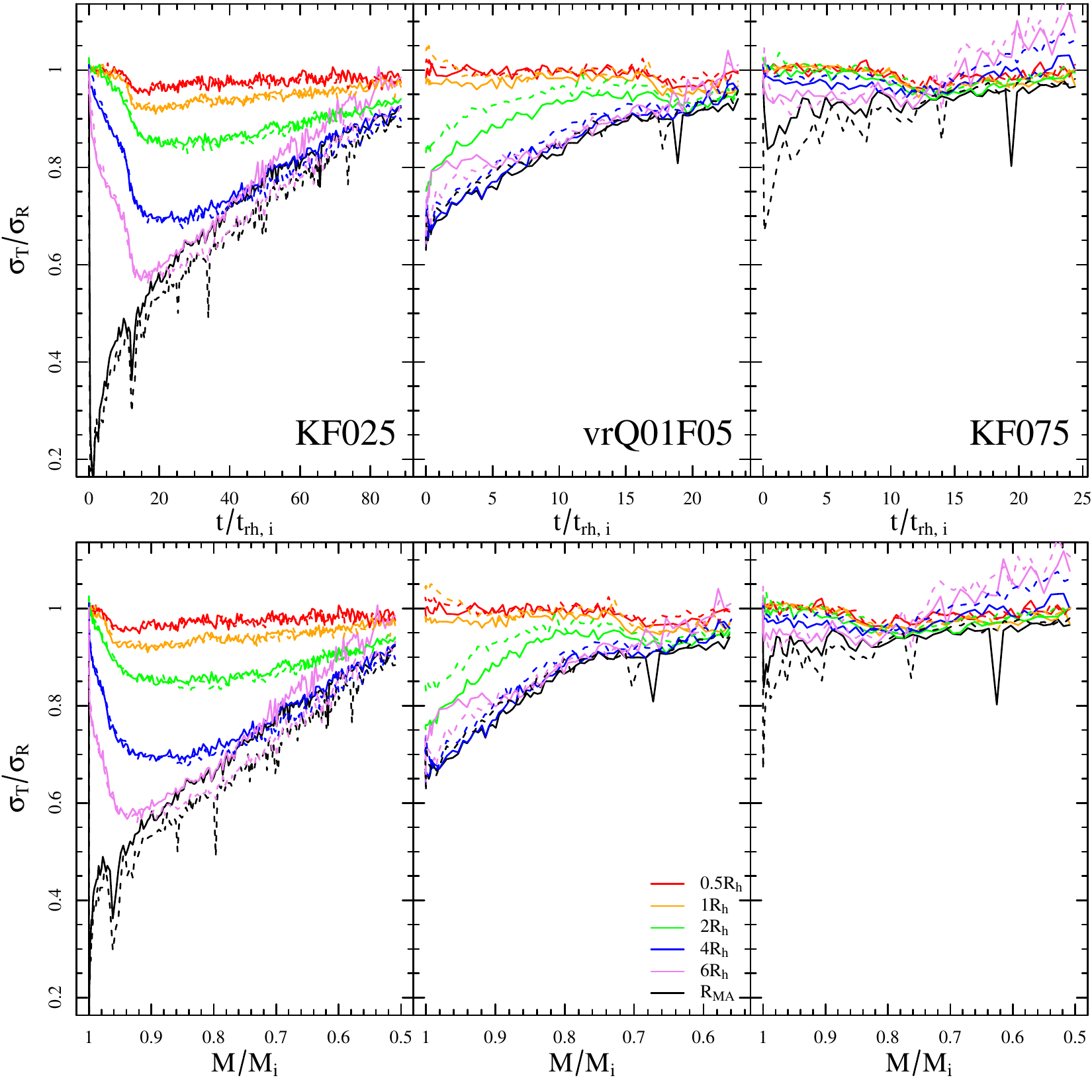}
\caption{Evolution of the anisotropy at multiples of $\Rh$ for a few representative models.  The top panels show the evolution as a function of time, and the bottom panels as a function of fraction initial mass remaining in the cluster.  Solid lines show the profiles calculated, as in previous figures, as projected along a line of sight perpendicular to the orbital plane; dashed lines show the profiles calculated as projected along a line of sight parallel to the orbital plane of the cluster.}
\label{fig:multRh}
\end{minipage}
\end{figure*}

Figure \ref{fig:multRh} shows the evolution of $\stsr$ at a few different distances
from the cluster centre for some of the models investigated in this
paper. These figures provide a more comprehensive picture of the radial dependence of the
anisotropy and its time evolution. 
A comparison of the models in Figure \ref{fig:multRh} further
illustrates how the outer region kinematics may provide key fingerprints
of a cluster evolution and initial structural properties.
Specifically, as already pointed out above and in Section
\ref{sec:sec31}, initially underfilling systems and systems undergoing
an early violent relaxation phase should be characterized by a strong
radial anisotropy in their outer regions unless they are in the advanced
stages of their evolution and have lost a significant fraction of their initial mass. Finally, in Figure \ref{fig:multRh} we also show the anisotropy measured along two different lines of sight and find no significant differences depending on the line of sight.

\section{Conclusions}
In this paper we have presented the results of an extensive survey of
\Nbody simulations aimed at exploring the long-term evolution of the velocity
anisotropy in tidally limited star clusters. We have studied the evolution of
three sets of models characterized by different initial structural and
kinematical properties: isotropic King models \citep{king1966},
rotating systems described by the models introduced by
\citet{varri2012}, and systems starting with a core-halo structure and
kinematical properties imprinted by an early phase of violent
relaxation in an external tidal field. 

We have explored in particular the dependence of the evolution of the
anisotropy and its radial variation within a cluster on
the initial filling factor (defined as the ratio of the cluster
half-mass to Jacobi radius, $\rhrj$) and have shown that, in all three sets of models, this is a key
parameter in determining the degree of radial anisotropy that can
develop in the intermediate and outer parts of a star cluster. 
Indeed, while the detailed evolution of the profile of the anisotropy as 
a function of radius show some differences in the three sets of simulations explored in this study, a global characteristic behavior clearly emerges.    

First of all, clusters with smaller
initial values of $\rhrj$ develop a stronger radial anisotropy.  
In addition, while the strength of the maximum radial anisotropy developed during a
cluster evolution strictly depends on the filling factor and the dynamical
phase, its location within a cluster is approximately constant and
falls in the range 0.2-0.4 $\rj$ during most of the cluster
evolution. 

Systems starting from isotropic, underfilling initial conditions during 
their evolution are typically characterized by an isotropic core
followed by a region with an increasing radial anisotropy, determined by the build-up of radial orbits in the outer parts of the cluster, reaching a
maximum between 0.2 and 0.4 $\rj$ and then decreasing again  and
becoming isotropic or mildly tangentially anisotropic in the cluster
outermost regions. 

As a cluster evolves and  starts to lose mass, the
growth of the radial anisotropy eventually stops and the systems evolves again
toward an isotropic velocity distribution at all distances from the
cluster centre. However such a global isotropic velocity distribution is
expected only for systems 
in the advanced stages of their evolution after they have lost a
significant fraction of their initial mass (we emphasize that we refer here to long-term mass loss due to two-body relaxation not to the early mass loss due to stellar evolution). Smaller
initial values of $\rhrj$ lead to stronger radial anisotropy and
require a larger amount of mass loss before their velocity
distribution becomes isotropic again. 

On the other hand, it is crucial to recognize that initially isotropic, tidally filling systems
never develop a significant radial anisotropy.

Interestingly, the behavior described above apply also to stellar systems starting from initial conditions characterized by moderate rotation and mild deviations from isotropy in the velocity space. 

As for star clusters undergoing an early phase of violent 
relaxation in an external tidal field, we show that, as previously found in \citet{vesperini2014}, these systems quickly develop during this phase an anisotropy profile similar to that described above (an isotropic core followed by regions with an increasing radial anisotropy reaching its maximum at $\sim$ 0.2-0.4 $\rj$ and then by isotropic or mildly tangentially anisotropic outermost regions).
During their long-term evolution these systems become more isotropic
but the anisotropy produced in their early evolution is, also in this
case, erased only after they have lost a significant fraction of their
initial mass.

The results of our simulations clearly show that information on the
initial structural properties and dynamical history should be
imprinted in the strength of the anisotropy in the intermediate and
outer regions of stellar systems. Provided that a star cluster has not
yet experienced severe mass loss, such information should be
accessible for many Galactic globular clusters. 

Several studies attempting to reconstruct the initial structural
properties of a number of Galactic globular clusters have suggested
that these systems might be initially characterized by small values of
$\rhrj$ \citep[see e.g.][]{giersz2011, gieles2011,pijloo2015} and that many of
them might not have lost a large fraction of their mass during their
long-term evolution. A strong
radial anisotropy in the 
intermediate/outer regions of these clusters is the expected
kinematical  fingerprint of the suggested initial conditions and
dynamical phase of these clusters. 

In addition to the obvious need to extend kinematical studies to a larger number of clusters, we emphasize the importance of extending the characterization of the degree of anisotropy to include star cluster intermediate and outer regions where, as shown by our results, the
kinematical fingerprints of different initial conditions and dynamical
history are expected to be stronger.

 \section*{Acknowledgments}
This research was supported in part by Lilly Endowment, Inc., through its support for the Indiana University Pervasive Technology Institute, and in part by the Indiana METACyt Initiative. The Indiana METACyt Initiative at IU is also supported in part by Lilly Endowment, Inc.  ALV acknowledges support from the Royal Commission for the Exhibition of 1851 and The Gruber Foundation in form of Research Fellowships.
\bibliographystyle{mn2e}
\bibliography{references}

\end{document}